\documentclass[
pre,citeautoscript,nofootinbib]{revtex4}

\usepackage{graphicx}
\usepackage{bm}
\usepackage{amsmath}
\usepackage{hyperref}

\begin{document}
\title{Derivation of the universal decay cascade distribution}
\author{Vyacheslavs Kashcheyevs}

\affiliation{Faculty of Physics and
Mathematics, University of Latvia, Riga LV-1002, Latvia}


\begin{abstract}
A detailed derivation of the decay cascade probability distribution stated in Eqs.\ (4)--(6) and (11) of Phys.\ Rev.\ Lett.\ \textbf{104}, 186805 (2010) [arXiv:0901.4102] by Kashcheyevs and Kaestner is provided. Recurrence relations are solved explicitly and connections between solutions in different limits are demonstrated.
\end{abstract}

\maketitle

\nocite{kaestner2010a}
\section{Mathematical definition of the decay cascade model~\cite{kaestner2010a}}

\nocite{kaestner2010a}
The probability distribution $P_n(t)$ is governed by the equation
\begin{align} \label{eq:mastereqs}
  \frac{d P_n(t)}{d t} & = -\Gamma_n(t) P_n(t) + \Gamma_{n+1}(t) P_{n+1}(t) \,  \tag{KK-1}
\end{align}
for $n \ge 0$ and with $\Gamma_0 \equiv 0$.
Normalization and initial conditions are
\begin{align}
  & \sum_{n=0}^{N} P_n(t)  =1  \, ,\\
  & P_n(t_0)  = \begin{cases} 
  1 , & n=N \\
  0, & n \not = N
  \end{cases} \, . \label{KK-2} \tag{KK-2}
\end{align}

Equation \eqref{eq:mastereqs} is a general kinetic equation for a birth-death Markov process \cite{Gardiner} for time- and population-size-dependent rates. Here we will  be interested in the asymptotic values of $P_n$ as the transition rates $\Gamma_n$ gradually decrease to zero as function of time.

The original publication  \cite{kaestner2010a} discussing solutions to the above model in the context of dynamic quantum dot initialization is denoted KK; equations marked here as KK-1, KK-2 etc.\ match the equations in KK with the corresponding numbers.

\section{Implicit exact solution}
In KK, the following exact iterative soltion, valid for $t>t_0$, is presented
\begin{align} \label{eq:ingralform} \tag{KK-3}
  P_n(t) &  = \int_{t_0}^{t} \! e^{
  - \int_{t'}^t \Gamma_{n}(\tau) \, d \tau } \Gamma_{n+1}(t') \,  P_{n+1}(t')  \, dt' \, , \\ 
  P_{N+1}(t) &  = \delta(t-t_0)/ \Gamma_{N+1}(t_0) \, . \label{eq:formal}
\end{align}
Condition \eqref{eq:formal} is introduced formally, in order for the general formula \eqref{eq:ingralform} to accommodate the initial condition \eqref{KK-2}; the delta functions are regularised as $\int_{t_0}^{t} \delta (t-t_0)=1$ for $t>t_0$.

Equation \eqref{eq:ingralform} is just the standard solution of a single linear first order differential equation  \eqref{eq:mastereqs} for an unknown function $P_n(t)$ with $P_{n+1}(t)$ treated as known. One can verify that \eqref{eq:ingralform}  solves \eqref{eq:mastereqs} by direct substitution:
\begin{align}
\frac{d P_n(t)}{d t} & =  \nonumber
  \underbrace{e^{
  - \int_{t}^t \Gamma_{n}(\tau) \, d \tau }}_{=1} \Gamma_{n+1}(t) \,  P_{n+1}(t)  
  + \int_{t_0}^{t} \! \frac{d}{d t} \left [ e^{
  - \int_{t'}^t \Gamma_{n}(\tau) \, d \tau } \right ]\Gamma_{n+1}(t') \,  P_{n+1}(t')  \, dt' \\ \nonumber
  & = \Gamma_{n+1}(t) \,  P_{n+1}(t)  + \int_{t_0}^{t} \! e^{
  - \int_{t'}^t \Gamma_{n}(\tau) \, d \tau } \underbrace{\frac{d}{d t} \left [- \int_{t'}^t \Gamma_{n}(\tau) \, d \tau \right ]}_{=-\Gamma_{n}(t)}  \Gamma_{n+1}(t') \,  P_{n+1}(t')  \, dt'
\\ \nonumber
& = \Gamma_{n+1}(t) \,  P_{n+1}(t) -\Gamma_{n}(t) P_n(t) \, .
\end{align}

\section{Explicit solution for time-independent rate ratio $\Gamma_n(t)/\Gamma_{n-1}(t)=\text{const}$}
The first solution described in KK corresponds
to the case of 
\begin{align} \label{eq:constantratio}
 \Gamma_n(t) = \frac{X_n}{X_1} \Gamma_1(t) \, ,
\end{align}
with $X_n \equiv \exp \sum_{k=1}^{n} \delta_k$ being time-independent constants.

For the rates obeying the condition \eqref{eq:constantratio}, the general solution can be constructed in the following form~\footnote{This form was inspired by studying explicit solutions for $\Gamma_n(t) \sim e^{-t}$ and $N=1,2,3$  with the means of a computer algebra system \texttt{Mathematica}.}
\begin{align}
  P_n(t) & = \sum_{k=n}^{N} R_{nk} \,
     e^{-\int_{t_0}^{t} \Gamma_k(t') dt'} \, , \label{eq:fullgeneral} 
\end{align}
with constant coefficients $R_{nk}$ that need to be determined.

\subsection{Derivation of $R_{nk}$}
For $n=N$, the initial conditions $P_N(t_0)=1$, $P_{N+1}(t_0)=0$ and equation \eqref{eq:fullgeneral} for $P_N(t)$ give
\begin{align} \label{eq:RNN}
 P_N(t) & =  e^{-\int_{t_0}^{t} \Gamma_N(t') dt'} \Longrightarrow R_{NN}=1 \, .
\end{align}
The initial condition $P_n(t_0)=0$ for  $n<N$,  applied to \eqref{eq:fullgeneral}, implies:
\begin{align}
  \sum_{k=n}^{N} R_{nk} = 0 \, , \quad n < N . \label{eq:total}
\end{align}

For $n<N$ the substituting \eqref{eq:fullgeneral} into the differential equation \eqref{eq:fullgeneral} gives
\begin{align}
\frac{d P_n(t)}{d  t}  & =
- \sum_{k=n}^{N} R_{nk} \, \Gamma_k(t) \, e^{-\int_{t_0}^{t} \Gamma_k(t') dt'} \label{eq:lhs} \\
-\Gamma_n(t) P_n(t) + \Gamma_{n+1}(t) P_{n+1}(t) & = 
 -\Gamma_n(t) \sum_{k=n}^{N} R_{nk} \,
    e^{-\int_{t_0}^{t} \Gamma_k(t') dt'}  +
    \Gamma_{n+1}(t) \sum_{k=n+1}^{N} R_{n+1,k} \,
  e^{-\int_{t_0}^{t} \Gamma_{k}(t') dt'}  \label{eq:rhs}
 \end{align}
 
Now we invoke the condition \eqref{eq:constantratio} which allows us to equate 
the coefficients of  $e^{-\int_{t_0}^{t} \Gamma_k(t') dt'}$ between \eqref{eq:lhs} and \eqref{eq:rhs}:
\begin{align} \label{eq:reucrrence}
  X_k R_{nk} & = X_n R_{nk} +X_{n+1} R_{n+1,k} \, , \quad {n< k \leq N } \\
  R_{n-1,k} & =\frac{X_{n}}{X_k - X_{n-1}}  R_{n,k}    , \quad {n \leq k \leq N } \tag{\ref{eq:reucrrence}'}
\end{align}

Equations \eqref{eq:total} and \eqref{eq:reucrrence} give the sought-after recurrence relations:
\begin{align}
  R_{nk} & = R_{kk} \prod_{m=n+1}^{k} \frac{X_{m}}{X_k - X_{m-1}} \, , \quad {n< k \leq N }  \label{eq:DC-5} \tag{KK-5'} \\
  R_{kk} & = - \sum_{m=k+1}^N R_{km} \, . \quad {k < N } \label{eq:DC-6} \tag{KK-6'}  
\end{align}
Equations \eqref{eq:DC-5} and \eqref{eq:DC-6} together with \eqref{eq:RNN} are equivalent 
to Eqs.~(5) and (6) of KK with  $C_k \equiv R_{kk}$ and $Q_{kn} \equiv R_{kn}/R_{kk}$.

\subsection{Explicit solution of recurrence relations}
The recurrence relations \eqref{eq:DC-5} and \eqref{eq:DC-6} admit the following explicit solution:
\begin{align} \label{eq:expsol}
   R_{nk} = \prod\limits_{m=n+1}^{N} X_m {\prod\limits_{\substack{m=n\\m\not=k}}^{N}} \frac{1}{X_k-X_m} \, .
\end{align}
The solution \eqref{eq:expsol} can be obtained for finite $n$ and $k$ by means of computer algebra, and proven in general form by induction.

Since integration over time preserves the condition \eqref{eq:constantratio}, we can choose
\begin{align}
X_n = \int_{t_0}^{t} \Gamma_n(t)\,  dt 
\end{align}
and write down the solution explicitly:
\begin{align} \label{eq:explicit}
  P_n(t) & = \sum_{k=n}^{N} e^{-X_k} \prod\limits_{m=n+1}^{N} X_m {\prod\limits_{\substack{m=n\\m\not=k}}^{N}} \frac{1}{X_k-X_m} \, .
\end{align}

The solution \eqref{eq:explicit} agrees precisely with the solution for $N=3$ and $\Gamma_n = \text{const}$ obtained by Miyamoto~\emph{et al.}~\cite{miyamoto2008}. 

\section{Solution in the limit of time-scale separation between cascade steps}
A more general solution that does not rely on condition \eqref{eq:constantratio} is derived in KK in the limit of decay time-scale separation between consecutive steps of the cascade. 
This is motivated as follows:
$P_{n+1}(t)$ stops changing appreciably over a times scale on the order of $\Gamma_{n+1}^{-1}(t)$ during which $P_{n}(t)$ changes only due to 
probability flux from state $(n+1)$, with negligible decay down the cascade to state $(n-1)$. This condition corresponds to $\Gamma_{n+1}(t) \gg \Gamma_n(t)$ during the relevant time interval. 

The mathematical part of the derivation proceeds as follows:
\begin{enumerate}
 \item Summing equations \eqref{eq:ingralform} for all $dP_m/dt$ with $N \ge m>n$ gives
\begin{equation}
\Gamma_{n+1}(t) P_{n+1}(t) = -\frac{d}{dt} \sum_{m>n}\!P_{m}(t).  
\end{equation}

\item On the ``slow'' time-scale controlled by $\Gamma_n(t)$, the function $\sum_{m>n}\!P_{m}(t)$
is changing rapidly from $1$ to its asymptotic value $\sum_{m>n}\!P_{m}(t\to \infty)$
and hence can be approximated by a step function in time. ($t \to \infty$ corresponds to the time when
the decay transitions no longer take place).

\item The derivative of a step function is proportional to a delta function and thus the exact integral
\eqref{eq:ingralform} can be approximated as follows:
\begin{align}\label{eq:sequential}
  P_n(t) &  = \int_{t_0}^{t} \! e^{
  - \int_{t'}^t \Gamma_{n}(\tau) \, d \tau } 
  \underbrace{\Gamma_{n+1}(t') \,  P_{n+1}(t')}_{\approx  \delta(t_0\!-\!t') [1-\sum_{m>n}\!P_{m}(t\to \infty)] }\, dt' \approx 
  \underbrace{e^{
  - \int_{t_0}^t \Gamma_{n}(\tau) \, d \tau }}_{e^{-X_n}}[1- \sum_{m>n}\!P_{m}(t\to \infty) ] \, .
 \end{align}
Equation \eqref{eq:sequential} essentially states that the decay of all previous states (higher than $n$) provides an initial condition for the decay of the $n$-th state. Based on \eqref{eq:sequential}, the condition on the final probabilities  $P_n(t\to \infty) \equiv P_n$ stated in KK is formulated:
\begin{align} \label{eq:condition}
 P_{n}\!=\!e^{-X_n}\left(1\!-\!\sum\nolimits_{m=n+1}^N P_m\right) \, . 
\end{align}

 \item Equation \eqref{eq:condition} is solved by expressing $P_{n-1}$ in terms of $P_n$. Using 
 \eqref{eq:condition} for $P_n$ we can express $\sum_{m>n} P_m  = 1- e^{X_n} P_n$ and 
 \begin{align}
 \sum_{m>n\!-\!1} P_m & = 1+(1-e^{X_n}) P_n  \, .\label{eq:sum}
  \end{align}
  Substituting the sum \eqref{eq:sum} into \eqref{eq:condition} for $P_{n-1}$, we get the derised recurrence relation
  \begin{align}
   P_{n-1} & = e^{-X_{n-1}}
  (e^{X_n} -1) P_n \, . \label{eq:Pnrecurrence}
 \end{align}
  Starting from $P_N = e^{-X_N}$ and iterating \eqref{eq:Pnrecurrence} gives
 \begin{align}
  \underbrace{
    \underbrace{
      \underbrace{e^{-X_N}}_{P_N} 
        \times \left (e^{X_{N}}-1 \right) e^{-X_{N-1}}}_{P_{N-1}} \times 
        \left (e^{X_{N-1}}-1 \right) e^{-X_{N-2}}}_{P_{N-2}} \times \ldots \, ,
 \end{align}
wherefrom the general form is easy to infer 
 \begin{align} \label{eq:DCcanonical}
    P_n = e^{-X_n} \prod_{m=n+1}^{N} \left ( 1- e^{-X_m} \right ) \, .\tag{KK-11}
 \end{align}
\end{enumerate}

Equation \eqref{eq:DCcanonical} has also been applied to
a generalised decay cascade scenario by Fricke~\emph{et al.} \cite{Fricke2013}.
They consider  the onset of decay steps at different times, $t_0< \ldots  < t^b_{n+1} < t^b_{n} < \ldots$, which in the notation of KK corresponds to
\begin{align}
   \Gamma_n(t) & = \tilde{\Gamma}_n(t) \Theta(t-t^b_n) \, ,
\end{align}
where $\Theta(x)$ is a unit step function and $\tilde{\Gamma}_n(t)$ are smooth functions that decay to zero as for $t\to \infty$.
The condition $X_n = \int_{t_0}^{\infty} \Gamma_n(t) = \int_{t^b_n}^{\infty} \tilde{\Gamma}_n(t) dt \gg X_{n-1}$
justifies the sequential cascade approximation \eqref{eq:sequential} and hence the probability distribution
\eqref{eq:DCcanonical}.


\end{document}